\documentclass[journal]{IEEEtran}

\ifCLASSINFOpdf
\else
   \usepackage[dvips]{graphicx}
\fi
\usepackage{url}
\usepackage{makecell}
\usepackage{xcolor}

 \usepackage{amssymb} 
\usepackage{graphicx}
\usepackage[numbers,sort&compress,square]{natbib}
\usepackage{svg}
\newsavebox{\measurebox}

\usepackage{algorithm}
\usepackage{algpseudocode}
\usepackage{dsfont}

\usepackage{amsmath}
\usepackage{multirow,multicol}
\usepackage{booktabs}
\usepackage{algorithm}
\usepackage{algpseudocode}
\usepackage[nolist]{acronym}
\usepackage{amsfonts}
\usepackage{xcolor}
\usepackage{graphicx}
\usepackage[normalem]{ulem}
\usepackage{footnote} 
\usepackage{subfigure} 
\usepackage{bm}
\usepackage{tabularx}

\newcommand{\fig}[1]{Fig. }
\newcommand{\tab}[1]{Table }
\newcommand{\Sec}[1]{Sec. }
\newcommand{\eq}[1]{Eq. }

\begin{document}
\newacro{QMUL}[QMUL]{Queen Mary University of London}
\newacro{ACE}[ACE]{Acoustic Characterization of Environments}
\newacro{STFT}[STFT]{Short-time Fourier Transform}
\newacro{RIR}[RIR]{Room Impulse Response}
\newacro{MAE}[MAE]{Mean Absolute Error}
\newacro{GLMB}[GLMB]{Generalized Labeled Multi-Bernoulli}
\newacro{TDOA}[TDoA]{Time Difference of Arrival}
\newacro{DoA}[DoA]{Direction of Arrival}
\newacro{ACC}[ACC]{\textit{Accuracy}}
\newacro{HRI}[HRI]{human robot interaction}
\newacro{AVMCF}[MCF]{Multi-channel Cost Function}
\newacro{AVGLF}[GLF]{Global Likelihood Function}
\newacro{AV-GLMB}[AV-GLMB]{Audio-Visual tracking with the Generalized Labelled Multi-Bernoulli}
\newacro{fps}[fps]{frame per second}
\newacro{DL}[DL]{Deep Learning}
\newacro{SP}[SP]{Signal Processing}
\newacro{SOTA}[SOTA]{state-of-the-art}
\newacro{MAE}[MAE]{Mean Absolute Error}
\newacro{KF}[KF]{Kalman Filter}
\newacro{STD}[STD]{STandard Deviation}
\newacro{CIL}[CIL]{{Class Incremental Learning}}
\newacro{SRPPHAT}[SRP-PHAT]{Steered-Response Power with Phase Transform}
\newacro{EKF}[EKF]{Extended Kalman Filter}
\newacro{DKF}[DKF]{Decentralized Kalman Filter}
\newacro{SSL}[SSL]{Sound Source Localization}
\newacro{PHD}[PHD]{Probability Hypothesis Density}
\newacro{FoV}[FoV]{Field-of-View}
\newacro{PF}[PF]{Particle Filter}
\newacro{AL}[AL]{Analytic Learning}
\newacro{PDF}[PDF]{Probability Density Function}
\newacro{GCC}[GCC]{Generalized Cross Correlation}
\newacro{GCF}[GCF]{Global Coherence Field}
\newacro{GCC-PHAT}[GCC-PHAT]{Generalized Cross Correlation with Phase Transform}
\newacro{STFT}[STFT]{Short-Time-Fourier-Transform}
\newacro{GLMB}[GLMB]{Generalized Labelled Multi-Bernoulli}
\newacro{MOT}[MOT]{Multiple Object Tracking}
\newacro{SOT}[SOT]{Single Object Tracking}
\newacro{RFS}[RFS]{Random Finite Set}
\newacro{FCN}[FCN]{Fully-connected Network}
\newacro{SMC}[SMC]{Sequential Monte Carlo}
\newacro{OSPA}[OSPA]{Optimal Sub-Pattern Assignment}
\newacro{MSE}[MSE]{\textit{Mean Square Error}}
\newacro{MAE}[MAE]{\textit{Mean Absolute Error}}
\newacro{MLP}[MLP]{Multiple Layer Perception}
\newacro{FE}[FE]{\textit{Feature Expansion}}
\newacro{SNR}[SNR]{Signal-to-Noise Ratio}

\newacro{FCM}[FCM]{\textit{Feature Correlation Matrix}}

\title{Analytic Class Incremental Learning for Sound Source Localization with Privacy Protection}

\author{Xinyuan Qian,~\IEEEmembership{Senior Member, IEEE},
Xianghu Yue, Jiadong Wang, Huiping Zhuang, Haizhou Li,~\IEEEmembership{Fellow, IEEE}
\thanks{Xinyuan Qian is with Department of Computer and Communication Engineering, University of Science and Technology Beijing, China (qianxy@ustb.edu.cn).}
\thanks{Xianghu Yue and Haizhou Li are with the National University of Singapore, 119077 Singapore (xianghu.yue@u.nus.edu, haizhou.li@nus.edu.sg).}
\thanks{Jiadong Wang is with MRI, Technical University of Munich, Germany (jiadong.wang@tum.de).}
\thanks{Huiping Zhuang is with Shien-Ming Wu School of Intelligent Engineering, South China University of Technology, China (hpzhuang@scut.edu.cn).}
\thanks{
% This research is supported by A*STAR under its AME Programmatic Funding Scheme (Project No. A18A2b0046) and its RIE2020 AME Programmatic Grant No. A1687b033.
The research is supported by National Natural Science Foundation of China (62306029, 62076024, 62006018, U22B2055), National Key Research and Development Program of China (2020AAA0109700), CCF-Tencent Rhino-Bird Open Research Fund,  A*STAR under its AME Programmatic Funding Scheme (Project No. A18A2b0046) and its RIE2020 AME Programmatic Grant No. A1687b033.
}
}

\markboth{Journal of \LaTeX\ Class Files, Vol. 14, No. 8, August 2015}
{Shell \MakeLowercase{\textit{et al.}}: Bare Demo of IEEEtran.cls for IEEE Journals}
\maketitle

\begin{abstract}
\ac{SSL} enabling technology for applications such as surveillance and robotics.
While traditional \ac{SP}-based \ac{SSL} methods provide analytic solutions under specific signal and noise assumptions,
recent \ac{DL}-based methods have significantly outperformed them.
However, their success depends on extensive training data and substantial computational resources.
Moreover, they often rely on large-scale annotated spatial data and may struggle when adapting to evolving sound classes.
To mitigate these challenges, we propose a novel \ac{CIL} approach, termed SSL-CIL, which avoids serious accuracy degradation due to \textit{catastrophic forgetting} by incrementally updating the DL-based SSL model through a closed-form analytic solution.
In particular, data privacy is ensured since the learning process does not revisit any historical data (exemplar-free), which is more suitable for smart home scenarios.
Empirical results in the public SSLR dataset demonstrate the superior performance of our proposal, achieving a localization accuracy of 90.9\%, surpassing other competitive methods.

\end{abstract}


\IEEEpeerreviewmaketitle

\section{Introduction}

In \ac{HRI}, robots typically employ \ac{SSL} is used to guide robotic attention.
Traditionally, \ac{SSL} is formulated within the \ac{SP} framework~\cite{knapp1976generalized,brandstein1997robust,schmidt1986multiple} to overcome robustness issues such as noise and reverberation~\cite{he2018deep}.
% Conventional \ac{SSL} techniques, which primarily leverage audio signals and formulate the challenge within the \ac{SP} framework \cite{knapp1976generalized,brandstein1997robust,schmidt1986multiple}, often struggle in acoustically complex environments characterized by noise and reverberation \cite{he2018deep}.
Recent studies have explored \ac{DL}-based methods \cite{chakrabarty2019multi,adavanne2018direction,he2018deep,pan2020multitones,qian2021multi} upon the availability of large-scale acoustic data with source location annotations.
These approaches have adopted distinct strategies: some map the location-based \ac{STFT} features to the \ac{DoA} of sounds \cite{chakrabarty2019multi,adavanne2018direction}, while others rely on cues derived from the \ac{GCC-PHAT} \cite{he2018deep,pan2020multitones,qian2021multi}.
Although the strong learning ability of \ac{DL} models ensures their remarkable success, their performance is penalized in scenarios of data scarcity or when data across different classes is presented in an incremental manner~\cite{qian2022audio} e.g., extensive data collection over time, labeled with various \ac{DoA}s, complements the diverse scenarios encountered in robotic audition.

% Urban life in the modern era has seen Sound Source Localization (SSL) technology take on an increasingly vital role. It plays a critical part not only in environmental monitoring, traffic management, and emergency response but also in enhancing the interaction experience of smart homes and Internet of Things (IoT) devices.

Emerging in recent years, \ac{CIL} is one of the novel solutions to mitigate the challenge of \textit{catastrophic forgetting}, specifically the inability to recall prior knowledge when encountering new data.
% In recent years, \ac{CIL}, an emerging machine learning method, offers innovative solutions to the challenge of  \textit{catastrophic forgetting} i.e.,  failure of recalling previously learned information when new data comes.
Unlike traditional batch learning methods, \ac{CIL} progressively assimilates new classes while retaining knowledge of prior ones. This capability allows for the continuous expansion of the knowledge base without necessitating retraining on the entire dataset.
% \ac{CIL} can update the model incrementally as new categories emerge without the need to retrain the entire dataset. 
This not only improves learning efficiency, but also reduces the demand for data storage and processing. 
The existing literature primarily focused on two strategies: regularization-based methods, such as less forgetting learning (LfL) \cite{jung2016less}, elastic weight consolidation (EWC) \cite{kirkpatrick2017overcoming} and learning without forgetting (LwF) \cite{li2017learning}; and replay-based techniques, including
iCaRL~\cite{rebuffi2017icarl},
SS-IL~\cite{Ahn2020SSILSS}
and reinforced memory
management (RMM) \cite{liu2021rmm}\cite{liu2020mnemonics}.
Specifically, the former strategy augments the loss function with novel regularization terms while the later one achieves \ac{SOTA} performance by integrating external memory modules or leveraging knowledge distillation to selectively revisit historical samples.
As an alternative, \ac{AL}-based~\cite{zhuang2022acil, zhuang2024, zhuang2023gkeal} methods exploit the iterative mechanism as its main cause and replace it with linear recursive tools. They achieve results comparable to those of replay-based techniques and exhibit robust performance.
% Nonetheless, the current AL-based methods may encounter the challenge of under-fitting due to their reliance on a single linear projection and the frozen backbone.

While previous \ac{CIL} methods have achieved promising results
in image classification~\cite{rebuffi2017icarl,li2017learning,Ahn2020SSILSS}, they have not been explored in spatial acoustic processing, especially the \ac{DoA} estimation of speakers.
Moreover, with the deployment of artificial intelligence in various fields, the protection of personal privacy issues has become increasingly prominent. 
Whereas traditional \ac{SSL} systems often require the collection and processing of large amounts of audio data, which may inadvertently infringe upon people's privacy. 
Thus, the challenge of effectively integrating \ac{CIL} strategy to \ac{SSL} tasks while ensuring effective privacy protection remains an open question.

To address the aforementioned issues, in this paper, we design a \ac{SSL} method that combines an analytic \ac{CIL} strategy to achieve rapid adaptation to new sound categories while protecting user privacy. Specifically, we propose a novel privacy protection mechanism that encrypts sensitive audio data during model training and inference processes, thereby preventing potential data leakage risks. 
% Through a series of experiments, we demonstrate that our proposed method maintains  the accuracy and robustness of \ac{SSL} while protecting privacy. 
Our research not only provides a new \ac{SSL} solution but also offers significant insights into how to balance performance and privacy protection in other \ac{HRI} tasks. The contributions are listed as follows.
\begin{enumerate}
    % \item While previous works target at improving the localization accuracy and robustness, we foresee their potential limitations in real-world \ac{HRI} scenarios and propose the first \ac{SSL} algorithm  under the \ac{CIL} setting
     % to locate the target in the spatial \ac{DoA} domain.
     \item To our best knowledge, we are the first \ac{SSL} proposal which locates the sounding object in the spatial \ac{DoA} domain under the \ac{CIL} setting.
    \item We propose the SSL-CIL algorithm, an analytic \ac{CIL} solution which avoids \textit{catastrophic forgetting} and potential exposure to data privacy without using previous knowledge when processing new \ac{DoA} data.
    \item Extensive experiments show the superior performance of our proposed SSL-CIL over the other competitors in terms of improved localization accuracy and robustness, less forgetting, and good privacy protection.
\end{enumerate}

\section{Proposed method}
\subsection{Problem formulation}
Let us denote the audio signals captured by a $M$-channel microphone array as ${\bf s}^{1:M}$, our objective is to estimate \ac{DoA} $\theta \in [1^\circ, 360^\circ]$  of the speaking person.
% As \ac{DoA}s are spatially continuous, instead of using the one-hot output coding, we use a Gaussian-like vector~\cite{he2018deep} to represent the \textit{posterior probability} {\textit{likelihoods}}, 
Given the spatial continuity of \ac{DoA}s, we replace the one-hot encoding with a Gaussian-like vector~\cite{he2018deep} to represent the posterior probability likelihoods:
% of a speaker presence in the direction of $\theta$, 
% \begin{equation}\label{eq:DoAencode}
%     {p}_t (\theta)= \exp \left (-\frac{|{ \theta}-\theta_t|^2}{\sigma^2_{\theta}}\right )
% \end{equation}
$ {p} (\theta)= \exp \left (-\frac{|{ \theta}-\dot{\theta}|^2}{\sigma^2_{\theta}}\right )$
where ${p}(\theta)$ is
% approximates a Gaussian function 
centered on the ground truth $\dot{\theta}$ with a standard deviation $\sigma_{\theta}$\footnote{The prefix $\frac{1}{\sqrt{2\pi \sigma_{\theta}}}$ of the Gaussian distribution is dropped, the same as~\cite{he2018deep}.}. 
Specifically, we design a network $\mathcal{F}$ to map the audio feature inputs to the speaker \ac{DoA} \textit{posterior probability},
% \begin{equation}\label{eq:prob}
    $\hat{p}(\theta) = \mathcal{F}(\mathbf{X}; {\bf \Omega})$
% \end{equation}
where  ${\bf \Omega}$ are the learnable parameters. Then, the  speaker \ac{DoA} is located with the highest probability value:
\begin{equation}
    \hat{\theta}=\arg \max_{\forall \theta} \ \hat{p}{(\theta)}
\end{equation}

Since \ac{DoA} data acquisition may encompass multiple phases, let us give the following definition.
A \textit{K}-phase CIL refers to a training approach where a network undergoes K phases, with each phase incorporating training data from distinct classes. Let ${\bf D}^\text{train}_k \sim \{ {\bf X}^\text{tr}_k, {\bf \Theta}^\text{tr}_k \}$ and ${\bf D}^\text{test}_k \sim \{ {\bf X}^\text{te}_k, {\bf \Theta}^\text{te}_k \}$ denote the training and testing datasets at phase $k=1,2,...,K$.
${\bf X}_k \in \mathbb{R}^{N_k\times d}$ and ${\bf \Theta}_k \in \mathbb{R}^{N_k\times 360}$ are the stacked $N_k$-sample audio feature inputs (with dimension $d$) and the posterior \ac{DoA} probability labels, respectively. In phase $k$, our proposed SSL-CIL aims to train a network given ${\bf D}^\text{train}_k$ and test them on all previous data ${\bf D}^\text{test}_{0:k}$ with seen classes, achieving less accuracy loss due to catastrophic forgetting.

\subsection{\ac{SSL} network}
\ac{GCC-PHAT}~\cite{knapp1976generalized} is chosen for its robustness to noise interference, achieved by utilizing solely the phase component of the cross spectrum while disregarding amplitude variations.
At microphone pair $n$, \ac{GCC-PHAT} is computed as:
\begin{equation}\label{eq:gccphat}
    g^n(\tau) = \sum_{f}\left(\frac{{\bf S}_{n_1}(f){\bf S}^{*}_{n_2}(f)}{|{\bf S}_{n_1}(f){\bf S}^{*}_{n_2}(f)|} e^{i \frac{2\pi f}{N_s} \tau}\right)
\end{equation}
where $i$, $f$ and $\tau$  denote the imaginary unit, frequency bin and time delay, $*$ is the complex conjugate. It exhibits the highest value in the time domain precisely at the actual time delay. 
% ~\footnote{All possible microphone pairs are used, thus $n\leq \frac{M(M-1)}{2}$.} 

We design a neural network~\cite{qian2021multi} to map  the input acoustic features to the output \ac{DoA} posterior probability:
% Specifically,  all \ac{GCC-PHAT}s are stacked to form the input audio feature  ${\bf X }$ .
\begin{equation}\label{eq:doa}
   \hat{\bf \Theta}=f_\text{softmax} \left ( \mathcal{F}_\text{mlp}({\bf X}, {\bf W}_\text{mlp}) {\bf W}_\text{fcn}\right )
\end{equation}
where ${\bf X \in \mathbb{R}^{N_k\times d}}$ is all the concatenated \ac{GCC-PHAT}s ($d=N_m\times \tau_r$ and $N_m=\frac{M(M-1)}{2}$ is the microphone pair number and $\tau_r$ is the possible GCC-PHAT range).
$\mathcal{F}_\text{mlp}$ represents a \ac{MLP} architecture comprising three hidden layers. Each is a \ac{FCN} layer with ReLU activation~\cite{nair2010rectified} and batch normalization~\cite{ioffe2015batch}.
${\bf W}_\text{mlp}$ and ${\bf W}_\text{fcn}$ denote the weights of \ac{MLP}  and the last \ac{FCN} classifier, respectively. 
$f_\text{softmax}$ is the softmax function.

\vspace{-2mm}
\subsection{Analytic re-alignment}
In the initial stage ($k=0$), we recalibrate the learning process of the network to align with the dynamics characteristic of the analytical approach.
Let us denote the extracted feature matrix ${\bf X}^\text{(mlp)}$ from the trained \ac{MLP} backbone as:
\begin{equation}\label{eq:feature}
    {\bf X}^\text{(mlp)}_{0} = \mathcal{F}_\text{mlp}({\bf X}^\text{tr}_{0}, {\bf W}_\text{mlp}) 
\end{equation}
where ${\bf X}^\text{(mlp)}_{0} \in \mathbb{R}^{N_k\times d_\text{(mlp)}}$. Rather than employing a  \ac{FCN} layer to directly project features onto the classification terminal, we implement a \ac{FE} process that strategically inserts an additional \ac{FCN} layer to achieve effective higher-dimensional feature mapping:
\begin{equation}\label{eq:expension}
\resizebox{0.85\linewidth}{!}{$
    {\bf X}^\text{(fe)}_{0}=f_\text{ReLU}\left ( \mathcal{F}_\text{mlp}({\bf X}^\text{tr}_{0}, {\bf W}_\text{MLP}) {\bf W}_\text{fe}\right ) = f_\text{ReLU} \left ( {\bf X}^\text{(mlp)}_{0}{\bf W}_\text{fe} \right ) $}
\end{equation}
where ${\bf X}^\text{(fe)}_{0} \in \mathbb{R}^{N_k\times d_\text{(fe)}}$ with $d_\text{(fe)}$ denotes the feature \textit{expansion size} ($d_\text{(fe)}>d_\text{(mlp)}$). The resulting increased parameters allow the \ac{AL}-based methods to obtain the maximum performance where every element in $d_\text{(fe)}$ is randomly sampled from a normal distribution~\cite{guo2001pseudoinverse,zhuang2021correlation}.
$f_\text{ReLU}$ denotes the ReLU activation function.
${\bf W}_\text{fe}$ is the \ac{FE} parameter matrix.

The expanded input acoustic features ${\bf X}^\text{(fe)}_{0}$ can be mapped to the \ac{DoA} labels through a linear regression process:
\begin{equation}\label{eq:linear}
    \mathop{\arg\min}\limits_{{\bf W}^{(0)}_\text{fcn}}
        \begin{Vmatrix}
 {\bf Y}^\text{tr}_{0}
-{\bf X}^\text{(fe)}_{0}
{\bf W}_\text{fcn}
\end{Vmatrix}^2_F
+\eta ||{\bf W}_\text{fcn}||^2_F
\end{equation}
where $||\cdot||^2_F$ denotes the Frobenius norm and $\eta$ is the regularization parameter. Then, the optimal solution (re-aligned weight) can be found as:
% and ${\bf W}_\text{(fcn)}$ indicates the final classifier layer.
\begin{equation}\label{eq:optimal}
    \hat{\bf W}^{(0)}_\text{fcn}=\left ( {\bf X}^{\text{(fe)}\intercal}_{0} {\bf X}^\text{(fe)}_{0}+\eta{\bf I}\right )^{-1}{\bf X}^{\text{(fe)}\intercal}_{0}{\bf \Theta}^\text{tr} _{0}
\end{equation}
where $(\cdot)^\intercal$ denotes transpose.

\begin{algorithm}[!tb]
\caption{The procedure of SSL-CIL}\label{alg:code}
\begin{algorithmic}
% \State 1. {\bf \ac{SSL} backbone training}: Train the \ac{SSL} backbone via BP on the base classes.
\Require Training data ${\bf D}^\text{train}_{0:k} \sim \{ {\bf X}^\text{tr}_{0:k}, {\bf \Theta}^\text{tr}_{0:k} \}$, regularization weight $\eta$.
\State initialize $k=0$
% \Ensure $y = x^n$
\State
(1) {\bf Analytic re-alignment:}
\State \ \ \ (i) obtain the expanded feature matrix ${\bf X}_{0}^\text{(fe)}$ with 
Eq.~\ref{eq:expension}
\State \ \ \ (ii) obtain the re-aligned weight $\hat{\bf W}^{(0)}_\text{fcn}$ with Eq.~\ref{eq:optimal} 
\State \ \ \ (iii) save the FCM matrix $\bf{R}_{0}$
\State 
(2) {\bf Class incremental learning:}
\For{$k$ from 1 to K   (with ${\bf D}^\text{train}_k$, $\hat{\bf W}^{(k-1)}_\text{fcn}$ and $\bf{R}_{k-1}$ )}
    \State (i) obtain the feature matrix ${\bf X}_{k}^\text{(fe)}$ with Eq.~\ref{eq:expensionk}
    \State (ii) update the FCM matrix $\bf{R}_{k}$ using Eq.~\ref{eq:FCMupdate}
    \State (iii) update the weight matrix $\hat{\bf W}^{(k)}_\text{fcn}$ using Eq.~\ref{eq:FCNweight}

\EndFor
\end{algorithmic}
\end{algorithm}

\vspace{-3mm}
\subsection{Class incremental learning}
After the analytic re-alignment, the SSL-CIL problem can be solved in an analytic manner.
Given the training data ${\bf D}^\text{train}_0$,...,${\bf D}^\text{train}_{k-1}$, the linear regression procedure can be formulated as:
\begin{equation}\label{eq:ICL}
% \argmin 
\mathop{\arg\min}\limits_{{\bf W}^{(k-1)}_\text{fcn}}
    \begin{Vmatrix}
 {\bf Y}^{tr}_{0:k-1}
-
{\bf X}^\text{(fe)}_{0:k-1}
{\bf W}^{(k-1)}_\text{fcn}
\end{Vmatrix}^2_F
+\eta ||{\bf W}^{(k-1)}_\text{fcn}||^2_F
\end{equation}
where 
\begin{equation}
\resizebox{0.83\linewidth}{!}{$
  {\bf Y}^{tr}_{0:k-1} =
  \begin{bmatrix}
 {\bf \Theta}^{tr}_0 & {\bf 0} & \hdots &  {\bf 0}  \\ 
 {\bf 0} &  {\bf \Theta}^{tr}_1  & \hdots  &  {\bf 0} \\ 
 & \vdots & \vdots &   \\ 
{\bf 0}& {\bf 0} & ... & {\bf \Theta}^{tr}_{k-1}  \\ 
\end{bmatrix}, \ 
{\bf X}^\text{(fe)}_{0:k-1}=
\begin{bmatrix}
 {\bf X}^\text{(fe)}_0\\ 
   {\bf X}^\text{(fe)}_1\\ 
 \vdots \\ 
 {\bf X}^\text{(fe)}_{k-1} \\ 
\end{bmatrix} $}
\end{equation}
where 
\begin{equation}\label{eq:expensionk}
    {\bf X}^\text{(fe)}_i=f_\text{ReLU}\left ( \mathcal{F}_\text{mlp}({\bf X}_i, {\bf W}_\text{MLP}) {\bf W}_\text{fe} \right ) 
    % = f_\text{ReLU} \left ( {\bf X}^\text{(mlp)}{\bf W}_\text{fe} \right )
\end{equation}

We give an LS-based analytical solution to Eq.~\ref{eq:ICL} for joint learning on ${\bf D}^\text{train}_{0:k-1}$, which can be written as:
\begin{equation}\label{eq:FCN}
\resizebox{0.85\linewidth}{!}{$
    \hat{\bf W}^{(k-1)}_\text{fcn}=\left ( \sum^{k-1}_{i=0} {\bf X}^{\text{(fe)}\intercal}_{i}  {\bf X}^\text{(fe)}_{i} +\eta {\bf I} \right )^{-1} 
    \left [   {\bf X}^{\text{(fe)}\intercal}_0{\bf \Theta}^\text{tr}_0 \hdots {\bf X}^{\text{(fe)}\intercal}_{k-1}{\bf \Theta}^\text{tr}_{k-1}  \right ] $}
\end{equation}
where $\hat{\bf W}^{(k-1)}_\text{fcn} \in \mathbb{R}^{{d_\text{(fe)}}\times360}$.

The goal of our proposed SSL-CIL is to provide an analytic solution at phase $k$ given $\hat{\bf W}^{(k-1)}_\text{fcn}$ and ${\bf D}^\text{train}_k$ without using any previous samples from ${\bf D}^\text{train}_{0:k-1}$. This is equivalent to obtain $\hat{\bf W}^{(k)}_\text{fcn}$ recursively based on $\hat{\bf W}^{(k-1)}_\text{fcn}$, data ${\bf X}^\text{(fe)}_k$ and ${\bf \Theta}^\text{tr}_k$ at current learning phase. Let us define a \ac{FCM} at phase $k-1$ as:
\begin{equation}\label{eq:FCM}
    \bf{R}_{k-1}=\left (\sum^{k-1}_{i=0} {\bf X}^{(fe)\intercal}_i {\bf X}^{(fe)}_i+\eta{\bf I} \right )^{-1}
\end{equation}
\textbf{Theorem 1}
Then the FCN weight defined in Eq.~\ref{eq:FCN} can be recursively obtained as:
\begin{equation}\label{eq:FCNweight}
\resizebox{0.85\linewidth}{!}{$
    \hat{\bf W}^{(k)}_\text{fcn}=\left [ \hat{\bf W}^{(k-1)}_\text{fcn} -{\bf R}_k {\bf X}^{\text{(fe)}\intercal}_k {\bf X}^\text{(fe)}_k\hat{\bf W}^{(k-1)}_\text{fcn} \ \  {\bf R}_k{\bf X}^{\text{(fe)}\intercal}_k {\bf \Theta}^\text{tr}_{k} \right ] $}
\end{equation}
where the FCM is also recursively updated as:
\begin{equation}\label{eq:FCMupdate}
\resizebox{0.85\linewidth}{!}{$
     \bf{R}_{k}= \bf{R}_{k-1}- \bf{R}_{k-1}{\bf X}^{(fe)\intercal}_k \left ({\bf I}+{\bf X}^{(fe)}_k {\bf R}_{k-1}{\bf X}^{(fe)\intercal}_k \right ) {\bf X}^{(fe)}_k {\bf R}_{k-1} $}
\end{equation}
Please see the \textit{Supplementary material} for \textit{Proof}.

To be noted, as indicated by Eq.~\ref{eq:FCNweight}, our proposed SSL-CIL method can recursively update the \ac{FCN} parameter matrix without using any historical data (exemplar-free), thus the data privacy is guaranteed.

\section{Experiments}
\noindent \textbf{Datasets.}
% ~\cite{he2018deep}.
% which is a valuable resource for research.
We  select the  SSLR dataset~\cite{SSRL} for experiments.
It was created using a real-world setup where four microphones are mounted on the head of the Peper robot, arranged in a 5.8 × 6.9 cm rectangular configuration, all of which are cardioid and front-facing in their directional orientation.

The audio sequences are captured at a sampling rate of 48 kHz, from one to two simultaneous speakers, complemented by precise 3D spatial annotations. 
% It is characterized by high-quality 4-channel audio ($M=4$) captured at a sampling rate of 48 kHz. 
The dataset is meticulously divided into three subsets: train-Loudspeaker, test-Human, and test-Loudspeaker.

\noindent \textbf{Settings.}
The \ac{GCC-PHAT} is calculated for each 170-ms segment with delay lags from -25 to 25, yielding 51 coefficients per microphone pair ($\tau_r=51$), consistent with the approach in \cite{he2018deep}. 
Given 6 microphone pairs ($N_m=6$), the total aggregate of GCC-PHAT coefficients obtained is 306 ($d=306$).
All models are trained for 30 epochs with a batch size of 256.
We use Adam~\cite{kingma2014adam} optimizer to train the model and set the learning rate and weight decay of 1e-3 and 1e-4, respectively.
We use \ac{MSE} loss to optimize the network.
For the replay-based methods, the memory size is set to 3600.
% ~\qian{can we add a ref? }.
For the incremental setting, we evenly divide the 360 classes into 10 phases ($K=10$), each of which contains 36 classes.
 
\noindent \textbf{Evaluation Metrics.}
% We adopt the \ac{MSE} loss for posterior probability-based coding.
We average the \textit{Mean Absolute Error} (MAE) and \textit{Accuracy} (ACC)  of the network incrementally trained at phase $k$ by testing it on ${\bf D}^\text{test}_{0:k}$ as the metrics. The ACC tolerance is set to $5^\circ$, the same as in~\cite{he2018deep,qian2021multi}, which allows the error variance to be in the range of [-5, 5] degrees.

% The final MAE and ACC are averaged MAE and ACC of the network incrementally trained at phase $k$ by testing it on ${\bf D}^\text{test}_{0:k}$.
% The symbols '$\uparrow$' and '$\downarrow$' indicate the desired direction of improvement.

\noindent \textbf{Baselines.} 
We compare the proposed SSL-CIL method with seven baselines, including 
1) \textit{SRP-PHAT}~\cite{brandstein1997robust}:  a \ac{SP}-based approach  which accumulates \ac{GCC-PHAT}s from multiple microphone pairs to determine the sound location; 
2) \textit{MLP-GCC}~\cite{he2018deep}: a conventional fully-supervised \ac{DL} method  which directly maps \ac{GCC-PHAT} to \ac{DoA}; and other five baselines  under \ac{CIL} settings, including
3) \textit{Joint training} (upperbound): it is trained using data from all previous tasks; 4) \textit{Fine-tune} (lowerbound): it initializes the model with parameters from the prior training step and re-trains it on current data without preventing catastrophic forgetting;
5) \textit{LwF}~\cite{li2017learning}: an exemplar-free method that preserves outputs of previous examples to reduce the forgetting of the old task and act as a regularizer for the new task;
6) \textit{SS-IL}~\cite{Ahn2020SSILSS}: a knowledge distillation network specialized for task-wise learning, utilizing a separated softmax output layer and task-wise knowledge distillation and
7) \textit{iCaRL}~\cite{rebuffi2017icarl}: it seamlessly incorporates new classes by leveraging a continuous sequential data stream. 
% To be noted, only 

\noindent \textbf{Results.} 
Table~\ref{tab:results} lists the comparison results of our proposed SSL-CIL algorithm with the conventional \ac{SP}-based, \ac{DL}-based localization methods and those \ac{CIL} algorithms. The best results in each column under the \ac{CIL} setting  is marked with bold font (excluding the upperbound of joint training). From the results, we can see that our proposed SSL-CIL can achieve the best localization results (MAE of 5.04 and ACC of 90.9\%), significantly outperforming other \ac{CIL} approaches, such as LwF (w data privacy)~\cite{li2017learning} (MAE of 37.06$^\circ$ and ACC of 46.0\%), SS-IL (w/o data privacy)~\cite{Ahn2020SSILSS} (MAE of 20.27$^\circ$ and ACC of 67.4\%) and iCaRL (w/o data privacy)~\cite{rebuffi2017icarl} (MAE of 24.4$^\circ$ and ACC of 70.5\%). Moreover, its performance is also close to the joint training (upperbound) method and the conventional \ac{DL}-based MLP-GCC~\cite{he2018deep}, where their overall ACC are all above 90\%. 
% \textcolor{blue}{To be noted, for the Human test set, the iCaRL sometimes achieve . However, it is worse than SSL-CIL in a statistic manner.}

\begin{figure}[!htb]
\centering
\includegraphics[scale=0.5]{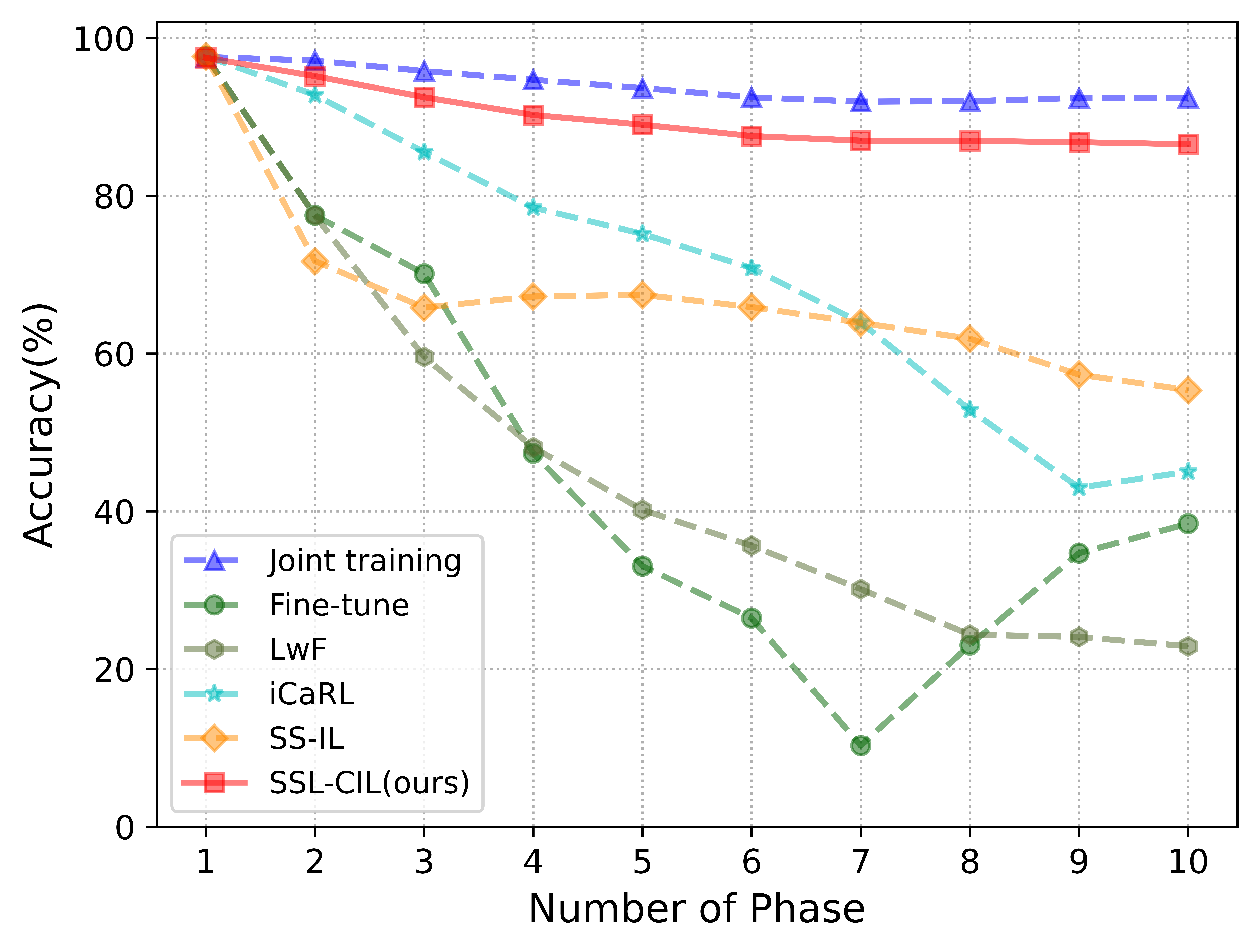}
\caption{The averaged \ac{DoA} estimation accuracy at various phases.
% of different methods (Note: joint training corresponds to the upperbound).
}
\label{fig:detection}
\end{figure}

\begin{table*}[!tb]
\centering
\caption{The speaker \ac{DoA} estimation results in MAE ($^\circ$) and ACC (\%) on the SSLR test set ($N$ denote the speaker count, with the audio frame count for each subset presented in parentheses.
% We reproduce the results of \cite{he2018deep} for comparison
`---': not applicable; $(\cdot)^{\clubsuit}$: upperbound; $(\cdot)^{\bigstar}$: lowerbound).}
\begin{tabular}{|l|c|c|c|c|c|c|c|c|c|c|c|c|c|c|}
 \toprule 
       &          & \multicolumn{4}{c|}{\textbf{Loudspeaker}}                                                   & \multicolumn{4}{c|}{\textbf{Human}}                                         & \multicolumn{2}{c|}{\multirow{2}{*}{\textbf{Overall}}}      
          \\ %\cmidrule{2-9} 
  \textbf{Model}  & \textbf{Privacy} & \multicolumn{2}{c}{\textbf{N=1   (178k)}}                                  & \multicolumn{2}{c|}{\textbf{N=2   (29k)}}                                   & \multicolumn{2}{c}{\textbf{N=1   (788)}}                                   & \multicolumn{2}{c|}{\textbf{N=2   (141) }}                                   & \multicolumn{2}{c|}{}            \\
       %                                  \cmidrule{2-11} 
                                        
       &  & \multicolumn{1}{l|}{MAE ($\downarrow$)}             & \multicolumn{1}{l|}{ACC ($\uparrow$)}             & \multicolumn{1}{l|}{MAE ($\downarrow$)}             & \multicolumn{1}{l|}{ACC ($\uparrow$)}             & \multicolumn{1}{l|}{MAE ($\downarrow$)}             & \multicolumn{1}{l|}{ACC ($\uparrow$)}             & \multicolumn{1}{l|}{MAE ($\downarrow$)}             & \multicolumn{1}{l|}{ACC ($\uparrow$)}             & \multicolumn{1}{l|}{MAE ($\downarrow$)}             & \multicolumn{1}{l|}{ACC ($\uparrow$)}          \\ \toprule  
     
\multicolumn{1}{|l|}{SRP-PHAT~\cite{brandstein1997robust}    }                &    ---                  & {19.00}         & {82.0}          & {36.95}         & {50.0}          & {2.62}          & {93.0}          & {20.90}         & {56.0}            & {21.44}         & {78.0}    \\  
\multicolumn{1}{|l|}{MLP-GCC~\cite{he2018deep}}          &                 ---                                & { 4.06}             & {94.9} & {8.10}          & {71.5}          & {4.75}          & {95.1}          & {5.98}          & {75.5}                 & {4.63}          & 91.6  \\  \midrule
\multicolumn{1}{|l|}{Joint training$^{\clubsuit}$}        &  $\times$      & 2.86 &  96.1 & 5.96 & 76.5  & 3.02 & 96.9 & 4.76  & 79.1 &  3.19   & 94.0    \\ 
\multicolumn{1}{|l|}{Fine-tune$^{\bigstar}$}          &   $\checkmark$                                   & 49.02  & 47.3 & 50.89  & 33.7 & 63.49 &  39.3  & 11.62  & 46.1  & 49.28 & 45.8   \\  
\multicolumn{1}{|l|}{LwF~\cite{li2017learning}}          &             $\checkmark$    &  36.90 & 47.2   &  38.56  & 34.8  & 33.44 & 53.7 & 36.89  & 13.5   & 37.06   &  46.0 \\
\multicolumn{1}{|l|}{SS-IL~\cite{Ahn2020SSILSS}}        &          $\times$    & 18.30 & 70.4 & 35.68  & 43.3 & 9.38  & 87.4  & 24.22 & 36.9 & 20.27  & 67.4    \\
\multicolumn{1}{|l|}{iCaRL~\cite{rebuffi2017icarl}}         &     $\times$  & 24.15  & 72.8  & 27.02 & 50.6  & {\bf 3.16} & 94.6  & {\bf 4.14} & {\bf 74.5}  & 24.40  & 70.5   \\   \midrule
 \multicolumn{1}{|l|}{SSL-CIL (ours)}                 &        $\checkmark$      & {\textbf{4.38}} & {\textbf{94.2}} & {\textbf{10.68}} & {\textbf{61.6}} & {3.35} & {\textbf{96.3}} &  {6.44}          & {{70.6}}  & {\textbf{5.04}} & {\textbf{90.9}}  \\ \bottomrule 
\end{tabular}
\label{tab:results}
\end{table*}

\begin{table*}[!htb]
\caption{Noise robustness of different \ac{CIL} methods under varying SNR conditions (from -20 to 10 dB).}
\centering
\begin{tabular}{|c|cc|cc|cc|cc|cc|cc|}
\toprule
\multirow{2}{*}{SNR} & \multicolumn{2}{c|}{Joint training$^{\clubsuit}$} & \multicolumn{2}{c|}{Fine-tune$^{\bigstar}$} & \multicolumn{2}{c|}{LwF~\cite{li2017learning}} & \multicolumn{2}{c|}{SS-IL~\cite{Ahn2020SSILSS}} & \multicolumn{2}{c|}{iCaRL~\cite{rebuffi2017icarl}}   & \multicolumn{2}{c|}{SSL-CIL (ours)} \\
 & MAE ($\downarrow$)  & ACC ($\uparrow$)         & MAE ($\downarrow$)   & ACC ($\uparrow$)        & MAE ($\downarrow$)  & ACC ($\uparrow$)         & MAE ($\downarrow$)  & ACC ($\uparrow$)       & MAE ($\downarrow$)  & ACC ($\uparrow$)        & MAE ($\downarrow$)   & ACC ($\uparrow$)     \\ \midrule
-20  & 40.24 & 13.3  & 63.07 & 8.4 & 49.58 & 10.6 & 48.08  & 10.1  & {\bf 45.98} & 11.1  & 46.43 & {\bf 11.5}  \\
-10 &  39.90 & 26.4 & 66.46 & 13.8  & 44.33 & 18.0             &  43.13  & 19.7  & 41.17  & 21.6 & {\bf 39.90} & {\bf 25.7}          \\
0    & 23.84 & 52.3 & 51.53 & 24.5  &  40.15 & 27.5         & 34.24 & 35.4  &  33.91 & 38.6 & {\bf 20.84} & {\bf 51.8}               \\
10  & 12.07 & 75.4 & 60.00 & 34.3 &  36.53 & 35.1  &        24.92 & 52.1 & 27.76 & 57.3  & {\bf 12.84}  & {\bf 72.7} \\ \midrule
Overall & 29.01 & 41.9 & 60.27 & 20.3 & 42.65 & 22.8 & 37.59 & 29.3 & 37.22 & 32.2 & {\bf 30.00} & {\bf 40.4} \\ \bottomrule 
\end{tabular}
\label{tab:noiseresult}
\end{table*}

\begin{table}[]
\centering
\caption{The impact of expansion size ($d_\text{(fe)}$
% of  ${\bf X}^\text{(fe)}_{(0)}$ in Eq.~\ref{eq:expension}
).}
\begin{tabular}{c|cc}
\toprule
Expansion size & MAE  & ACC   \\ \hline
1000           & 6.68 & 88.0 \\
5000           & 5.61 & 89.9 \\
10000          & 5.33 & 90.3 \\
15000          & 5.14 & 90.7 \\
20000          & {\bf 5.04} & {\bf 90.9} \\  \toprule
\end{tabular}
\label{tab:expansionsize}
% \vspace{-3mm}
\end{table}

\begin{table}[]
\centering
\caption{The impact of regularization factor ($\eta$).}
\begin{tabular}{c|cc}
\toprule
Regularization factor & MAE  & ACC   \\ \hline
1           & 5.58 & 90.1 \\
0.1         & {\bf 5.04} & {\bf 90.9} \\
0.01        & 5.43 & 86.3 \\ \toprule
\end{tabular}
\label{tab:regularization}
% \vspace{-3mm}
\end{table}

We also evaluate the noise robustness of different methods with the varying \ac{SNR}s ranging from -20 to 10 dB. The results are listed in Tab~\ref{tab:noiseresult} where we can see that our proposed SSL-CIL achieves the best results with the overall MAE of 30$^\circ$ and an ACC of 40. 4\%. It should be noted that, compared to joint training,  our performance only degrades slightly when the ACC of both methods is greater than 70\%, 50\%, 25 \% and 10\% under decreasing \ac{SNR}s.
% varies from 10 dB to -20 dB.
It also superiors the replay-based iCaRL method~\cite{rebuffi2017icarl} that obtains the overall MAE and ACC of 37.22$^\circ$ and 32.2\%, respectively.

Fig.~\ref{fig:detection} shows the averaged DoA estimation accuracy in various phases of different methods. It is shown that our proposed SSL-CIL method (red curve) obtains the best performance at each incremental step (excluding the upperbound), with less forgetting, better
accuracy, and data privacy.
In summary, our proposal has shown obvious superiority compared to others, which demonstrates its effectiveness.

\noindent \textbf{Ablation Study.}
To quantitatively evaluate the effect of the involved hyper-parameters, we conduct ablation experiments on the impact of expansion size $d_\text{(fe)}$ (in Eq.~\ref{eq:expension}) with the results given in Table~\ref{tab:expansionsize}. The FE process is justified by the fact that \ac{AL}-based methods need
more parameters to obtain the maximum performance.
This statement is validated, since both the MAE and ACC results improve when increasing the expansion size. Thus, we empirically set $d_\text{(fe)}$ at 20000.
% \qian{XXX}
We also ablate the regularization factor $\eta$ (Eq.~\ref{eq:linear} and Eq.~\ref{eq:ICL}) in Table~\ref{tab:regularization}. It is indicated that when we set $\eta$ to 0.1, the SSL-CIL algorithm achieves the best performance, that is, MAE of 5.04$^\circ$ and ACC of 90.9\%.
% \vspace{-2mm}
\section{Conclusion}

In this paper, we propose the SSL-CIL algorithm to tackle the challenging speaker \ac{DoA} estimation problem under the \ac{CIL} settings for the first time.
Specifically,  SSL-CIL provides an analytic solution that facilitates the incremental learning of new sound classes through a recursive manner.
It integrates two critical features: exemplar-free and data privacy protection to overcome key limitations in the \ac{CIL} field.
The experimental results in the public SSLR dataset demonstrate competitive performance over the other \ac{SOTA} methods, achieving MAE and ACC of 5.04$^\circ$ and 90.9\%, respectively.
% , which is almost close to the joint training (upperbound).
It also shows better results when tested under different \ac{SNR} conditions.
Future work may extend this study to the speaker tracking scenario.

\newpage
\bibliographystyle{IEEEtran}
\bibliography{main}

\newpage
\appendix
\section*{Proof of Theorem 1}
According to the Woodbury matrix identity, for any invertible square matrices ${\bf A}$ and ${\bf C}$, we have
\begin{align}
    ({\bf A} + {\bf UCV})^{-1} = {\bf A}^{-1} - {\bf A}^{-1} {\bf U} ({\bf C}^{-1}+{\bf VA}^{-1}{\bf U} ){\bf VA}^{-1}
\end{align}
Let ${\bf A} = {\bf R}_{k-1}^{-1}$, ${\bf U} = {\bf X}_k^{\text{(fe)}\intercal}$, ${\bf C} ={\bf I}$, ${\bf V} = {\bf X}_k^{\text{(fe)}}$. From the recursive \ac{FCM} update $ {\bf R}_{k} = \left ({\bf R}_{k-1}^{-1} + {\bf X}_k^{\text{(fe)}\intercal}  {\bf X}_k^{\text{(fe)}} \right )^{-1}$
% \begin{align}
% \label{eq_r_m_k}
%     {\bf R}_{k} = \left ({\bf R}_{k-1}^{-1} + {\bf X}_k^{\text{(fe)}\intercal}  {\bf X}_k^{\text{(fe)}} \right )^{-1},   
% \end{align}
we have
\begin{align}
\label{eq_R_update1}
    {\bf R}_{k} = {\bf R}_{k-1} - {\bf R}_{k-1}{\bf X}_k^{\text{(fe)}\intercal} \left ({\bf I}+{\bf X}_k^{\text{(fe)}}{\bf R}_{k-1}{\bf X}_k^{\text{(fe)}\intercal} \right ){\bf X}_k^{\text{(fe)}}{\bf R}_{k-1}
\end{align}
which completes the proof for the recursive formulation of \ac{FCM}.
To facilitate subsequent calculations,  let us define
\begin{align}
\label{eq_Q_k_1}
    {\bf Q}_{k-1} = {\bf X}_{0:k-1}^{\text{(fe)}\intercal} {\bf \Theta}^\text{tr}_{0:k-1}
\end{align}
According to Eq.~\ref{eq:FCN}, Eq.~\ref{eq:FCM} and Eq.~\ref{eq_R_update1}, we have
\begin{equation}
\begin{aligned}\label{eq:W}
    \hat{\bf W}^{(k)}_\text{fcn}={\bf R}_k \left [ {\bf Q}_{k-1} \ \ {\bf X}_{k}^{\text{(fe)}\intercal} {\bf \Theta}^\text{tr}_k \right ] \\
    =\left [{\bf R}_k {\bf Q}_{k-1} \ \ {\bf R}_k  {\bf X}_{k}^{\text{(fe)}\intercal} {\bf \Theta}^\text{tr}_k \right ]
    \end{aligned}
\end{equation}
where 
\begin{equation}
\begin{aligned} \label{eq:RQ}
    & {\bf R}_{k} {\bf Q}_{k-1}  \\ 
    &= {\bf R}_{k-1} {\bf Q}_{k-1} - {\bf R}_{k-1} {\bf X}_{k}^{\text{(fe)}\intercal} \left ({\bf X}_{k}^{\text{(fe)}} {\bf R}_{k-1} {\bf X}_{k}^{\text{(fe)}\intercal} + {\bf I} \right )^{-1} {\bf X}_{k}^{\text{(fe)}} {\bf R}_{k-1} {\bf Q}_{k-1}\\
    & = \hat{\bf W}^{(k-1)}_\text{fcn} - {\bf R}_{k-1} {\bf X}_{k}^{\text{(fe)}\intercal}\left ({\bf X}_{k}^{\text{(fe)} }{\bf R}_{k-1} {\bf X}_{k}^{\text{(fe)}\intercal} + {\bf I} \right )^{-1} {\bf X}_{k}^{\text{(fe)}} \hat{\bf W}^{(k-1)}_\text{fcn}
\end{aligned}
\end{equation}

Let ${\bf K}_k = \left ({\bf X}_{k}^{\text{(fe)}} {\bf R}_{k-1} {\bf X}_{k}^{\text{(fe)}\intercal} + {\bf I} \right )^{-1}$. Since 
\begin{align}
    {\bf I} = {\bf K}_k {\bf K}^{-1}_k = {\bf K}_k \left ({\bf X}_{k}^{\text{(fe)}} {\bf R}_{k-1} {\bf X}_{k}^{\text{(fe)}\intercal} + {\bf I} \right )
\end{align}
we have
${\bf K}_k = {\bf I} - {\bf K}_k {\bf X}_{k}^{\text{(fe)}} {\bf R}_{k-1} {\bf X}_{k}^{\text{(fe)}\intercal}$.
Therefore,         
\begin{align}\label{a}
    & {\bf R}_{k-1} {\bf X}_{k}^{\text{(fe)}\intercal} \left ({\bf X}_{k}^{\text{(fe)}} {\bf R}_{k-1} {\bf X}_{k}^{\text{(fe)}\intercal} +{\bf I} \right )^{-1}
    \notag \\
    &= {\bf R}_{k-1} {\bf X}_{k}^{\text{(fe)}\intercal} {\bf K}_k
    \notag \\
    &= {\bf R}_{k-1} {\bf X}_{k}^{\text{(fe)}\intercal}  \left ( {\bf I} - {\bf K}_k {\bf X}_{k}^{\text{(fe)}} {\bf R}_{k-1} {\bf X}_{k}^{\text{(fe)}\intercal}  \right )
    \notag \\
    &= \left ({\bf R}_{k-1} - {\bf R}_{k-1} {\bf X}_{k}^{\text{(fe)}\intercal} {\bf K}_k {\bf X}_{k}^{\text{(fe)}} {\bf R}_{k-1} \right ) {\bf X}_{k}^{\text{(fe)}\intercal}
    \notag \\
    &= {\bf R}_{k} {\bf X}_{k}^{\text{(fe)}\intercal}
\end{align}

Thus, Eq.~\ref{eq:RQ} can be reduced to:
\begin{align}
\label{eq_RQprime}
    {\bf R}_{k} {\bf Q}_{k-1} = \hat{\bf W}^{(k-1)}_\text{fcn} - {\bf R}_{k} {\bf X}_{k}^{\text{(fe)}\intercal} {\bf X}_{k}^{\text{(fe)}} \hat{\bf W}^{(k-1)}_\text{fcn}
\end{align}
By substituting Eq.~\ref{eq_RQprime} into Eq.~\ref{eq:W}, we can complete the proof of \textbf{Theorem 1},
\begin{align*}
    \hat{\bf W}^{(k)}_\text{fcn} = \left [ \hat{\bf W}^{(k-1)}_\text{fcn} - {\bf R}_{k} {\bf X}_{k}^{\text{(fe)}\intercal} {\bf X}_{k} \hat{\bf W}^{(k-1)}_\text{fcn}  \ \ {\bf R}_{k} {\bf X}_{k}^{\text{(fe)}\intercal} {\bf \Theta}^\text{tr}_k \right ]
\end{align*}
which proves the recursive calculation of $\hat{\bf W}^{(k)}_\text{fcn}$.

\end{document}